\documentstyle[twoside,fleqn,espcrc2,epsf]{article}
\newcommand{\bye}{\end{document}}

\newcommand{\be}{\begin{equation}}
\newcommand{\ee}{\end{equation}}
\newcommand{\beq}{\begin{equation}}
\newcommand{\eeq}{\end{equation}}
\newcommand{\bes}{\begin{eqnarray}}
\newcommand{\ees}{\end{eqnarray}}
\newcommand{\bma}{\left( \begin {array}}
\newcommand{\ema}{\end {array} \right)}
\newcommand{\xslash}[1]{\kern 0.2 em #1\kern -0.35em /}
\newcommand{\cslash}{\kern 0.1 em C_i \kern -0.75em / \kern 0.2 em }
\newcommand{\pslash}{\kern 0.2 em p\kern -0.45em /}
\newcommand{\sla}[1]{\kern 0.2 em #1\kern -0.45em /}
\newcommand{\prd}{ Phys.~Rev.\ D}
\newcommand{\npb}{Nucl.~Phys.\ B}
\newcommand{\prl}{Phys.~Rev.~Lett.\ }
\newcommand{\plb}{Phys.~Lett.\ B}

\newcommand{\cona}{Nucl.~Phys.\ B[Proc.~Suppl.]20 (1991) }
\newcommand{\conb}{Nucl.~Phys.\ B[Proc.~Suppl.]26 (1992) }
\newcommand{\conc}{Nucl.~Phys.\ B[Proc.~Suppl.]30 (1993) }
\newcommand{\ms}{\overline{M\!S}}
\newcommand{\qq}{q \bar q}
\newcommand{\ra}{\rightarrow}
\newcommand{\lra}{\longrightarrow}

\newcommand{\bc}{\begin{center}}
\newcommand{\ec}{\end{center}}
\newcommand{\bi}{\begin{itemize}}
\newcommand{\ei}{\end{itemize}}

\title{$\alpha_s$ from the Lattice Potential}

\author{\underline{K.~Schilling}\thanks{Work supported
        by DFG grant Schi 257/1-4 and EC project SC1*-CT91-0642.}
        and G.S.~Bali\vskip\baselineskip
        Department of Physics, University of Wuppertal,
        D-42097 Wuppertal, Germany}

\begin{document}


\begin{abstract}
We present an extensive
study on the direct determination
of  the running coupling $\alpha_s$ from  the static quark
antiquark force  at short distances, in quenched QCD. We find from
our high statistics potential analysis that
$\alpha_{q\bar q}$
exhibits two-loop asymptotic behaviour
for momenta as low as $.5$~GeV. As a result,
we determine the zero flavour $\Lambda$-parameter
to be $\Lambda^{(0)}_{\ms} =
.630(38)\sqrt{\sigma} = 293(18)^{+25}_{-63}$~MeV.
A rough estimate of full QCD effects leads to the
five  flavour value $\alpha_{\overline{M\!S}}(m_Z) = .102^{+6}_{-11}$.
A comparison  with other lattice  results is made.
\end{abstract}
\maketitle

\section{INTRODUCTION}
Considerable progress has been achieved over the past three years in the
determination of  the running coupling from SU(3) gauge
theory, {\it i.e.} quenched  QCD
on the lattice. Just like in real experiment,
this is not an easy task, since we are dealing with logarithmic effects.
Moreover,  asymptotic
scaling is substantially violated in the
$\beta$-window
presently accessible to lattice
simulations. This
is to say that
the perturbative two-loop-formula
does not yet relate the lattice spacing $a$ to  the bare
coupling, $g_0^2 = 6/\beta$, in terms of a {\it constant} lattice
scale $\Lambda _L$:
\be
\label{relaqqbar}
a\Lambda _L=
\exp\left(-\frac{1}{2b_0g_0^2}\right)(b_0g_0^2)^{-\frac{b_1}{2b_0^2}}\quad,
\ee
with
\be
b_0=\frac{11}{16\pi^2}\quad,\quad
b_1=  \frac{102}{\left(16\pi^2\right)^2}\quad .
\ee
Therefore this perturbative relation
in the bare lattice coupling scheme is not
well suited for a determination of
$\alpha_{s}$.
This  scheme
is also too far off
the continuum $\ms$-scheme
to rely on the perturbative recoupling~\cite{hasenfratz}
\be
\label{relalatt}
\alpha_{\ms} = \alpha_L + 5.88\alpha^2_L + {\cal O}(\alpha^3_L),\quad
\alpha_L = g_0^2/4\pi
\ee
at accessible lattice spacings.

Three major directions have been followed
in the past to tackle the problem:
(a) abandon
the bare lattice coupling for the sake of  ``improved'' coupling
schemes on the lattice, which are better suited to
renormalized perturbative expansions of short
distance operators~\cite{mackenzie90};
(b) determine $\alpha_{\qq}$ directly from the
static $q \bar q$-force,
on sufficiently large and fine lattices~\cite{michael,bsrun};
(c) contrive a scheme with a
volume
dependent coupling $g(L)$.
This   novel access to the problem,
put forward in
Ref.~\cite{luescher1},
is intriguing as it
lends itself to  the application of
finite-size matching
techniques.
\begin{figure}[htb]
\vskip -1cm
\epsfxsize=300pt
\epsfbox[60 210 660 520]{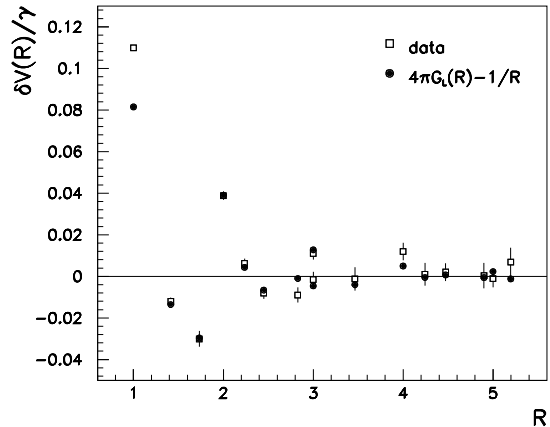}
\caption{Comparison between the oscillation of the
potential data (at $\beta = 6.4$)
around $V_{cont}$ and the expectation of Eq.~(8).\vskip -.3cm}
\label{fig:unfold}
\end{figure}

Method (c) has been covered during last year's
conference with application to the SU(2) case~\cite{luescher92}.
The outcome for SU(3) gauge theory is
presented in WOLFF's talk~\cite{wolff},
while the status of
method (a) has been reviewed in the contribution
of EL-KHADRA to this session~\cite{elkhadra}.

We will focus here on the direct
determination of $\alpha_{\qq}$ from the short distance force
and report on the
status of the  analysis of high precision potential data
in SU(3) gauge theory.
In view of the quality
of today's data, the analysis method~\cite{michael,protvino} is devised
to extract   $\alpha_{\qq}$ {\it locally} at short distances rather
than deriving it from  a global ansatz to the force -- with
short and long distance boundary conditions -- as proposed  in
Ref.~\cite{rebbibarkai}.

We perform the ``classical''  Creutz
experiment~\cite{creutz79} and extract the potential from the local masses
\be
V(\vec R) = \ln\frac{W(R,T)}{W(R,T+1)} \quad,\quad T\,\,\,\mbox{large}\quad,
\ee
using improved techniques for  suppression
of   excited state contributions
in order  to attain an early onset of the asymptotic  plateau in $T$,
as explained in Refs.~\cite{bspot,bsrun}.
\begin{table}
\begin{center}
\begin{tabular}{|c|c|c|c|c|c|}
\hline
$\beta = $&5.5 & 5.6 & 5.7 & 5.8 & 5.9\\
\hline
$V = 16^4$  &400 & 200 & 200 & 400 & 220  \\
\hline
\hline
$\beta = $ & 6.0 & 6.2 & 6.4 & 6.8  & \\
\hline
$V = 32^4$ & 185 & 200 & 252 & 110 &  \\
\hline
\end{tabular}
\end{center}
\caption{The number of independent configurations produced and
analysed  on  $16^4$ and $32^4$
lattices.}
\label{tab1}
\end{table}
On more than 2000 independent configurations
of $32^4$ and $16^4$ lattices (Tab.~\ref{tab1}),
measurements of on- and off-axis
potentials have been carried out for 72 and 36 different
$q\bar q$ separations
$\vec R$, respectively, with resolutions ranging from $.03$
to $.25$~fm.
UKQCD has  another set of
50 $36^4$ configurations at $\beta = 6.5$ with
mostly on-axis potentials~\cite{ukqcd} measured.
\begin{figure}[htb]
\vskip -0.5cm
\epsfxsize=300pt
\epsfbox[60 210 660 530]{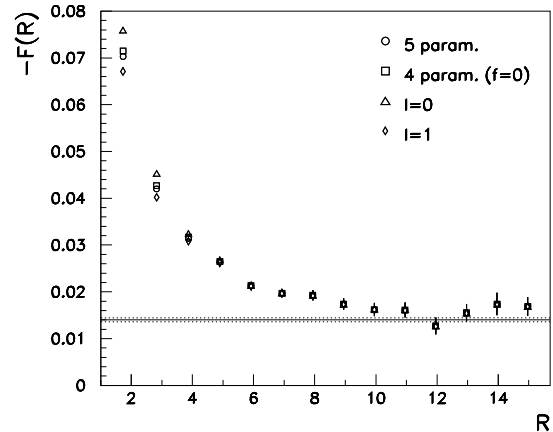}
\caption{The sensitivity of the on-axis force at $\beta = 6.4$ on
the parametrization of the potential.\vskip -.3cm}
\label{fig:sensitivity}
\end{figure}

The interquark
force $F$ is expected to behave perturbatively
at sufficiently short distances:
\be
 F(R) = - \frac{4}{3} \alpha_{q\bar q}(R)/R^2\quad .
 \ee
Perturbation theory relates
the  $\qq$- and the $\ms$- schemes
by means of a very  small first-order coefficient~\cite{billoire}:
\be
\label{smallcoeff}
\alpha_{\ms} = \alpha_{\qq} - .086\alpha^2_{\qq} + {\cal
O}(\alpha^3_{\qq})\quad .
\ee
This  connection lets  the $\qq$-scheme
appear as  a  very promising  candidate
for a ``perfect'' lattice renormalization scheme~\cite{mackenzie90}.

\section{UNFOLDING LATTICE ARTIFACTS}
At present, an  $\alpha_s$-analysis of the potential must rely on
 lattice data in the $R$-region from 2 to 7.
This necessitates  a  careful treatment of
lattice artifacts.

One expects~\cite{rebbilang} the continuum and lattice
potentials to be
related by a correction term
\be
\label{correct}
V(\vec{R})  = V_{cont}(R) - \delta V(\vec{R})\quad,
\ee
which should be proportional to the difference between
the lattice and continuum  propagators:
\be
\label{corr}
\delta V({\vec R}) = \gamma \left(4\pi G_L({\vec  R}) - 1/R\right)\quad.
\ee
As one can see from Fig.~\ref{fig:unfold} (circles),
this correction exhibits significant oscillations around zero for
$ R \leq 4$, with maximal deflection occurring
for the on-axis lattice potentials.
For
the  determination of  the parameter $\gamma$ in Eq.~(\ref{corr}),
we make use of two  parametrizations of the
continuum potential. We follow MICHAEL~\cite{michael} and choose as the first
form
\be
\label{potform}
V_{cont}(R)=V_0+KR-e/R+f/R^2\quad ,
\ee
while the second is the Cornell potential from Eq.~\ref{cornell}
($f = 0$).
We demonstrate
in Fig.~\ref{fig:unfold}, that this simple ansatz
works surprisingly well:
with the best fit value for $\gamma /e = l \approx .6$, which incidentally
comes out
independently
of $\beta$ and $f$,
the lattice potential data are seen to bounce around the  interpolating
curve in very much the same way as described by Eq.~(\ref{corr}).

We emphasize again, that at this stage we are merely concerned
with unfolding the lattice artifacts, and we better
make sure that the unfolding procedure
is sufficiently robust in the sense, that
the resulting  ``continuum'' force, $F$,
is not an artifact of the underlying interpolative
ansatz Eq.~(\ref{potform}).

To that end, let us first  define the force
by discrete differentiation according to
\be
\label{diff}
F(\bar R) = \\
\frac{V(\vec R_1) - V(\vec R_2)}{\Delta},\quad \Delta = R_2 - R_1.
\ee
This definition  carries an error ${\cal  O}\left((\Delta/\bar R)^2\right)$.

In order to reduce
this discretization error,
$\bar R$  should be properly placed within   the
interval $(R_1,R_2)$.  We have exploited the parametrization,
Eqs.~(\ref{correct},\ref{corr},\ref{potform}),
to compute this tangential point and estimate its error from the
uncertainties of
the fit parameters $l = \gamma/e$ and $f/e$. We assumed systematic
errors of $20 \%$ and $100 \%$ on these parameters, respectively.
The discretization effects, estimated in this manner,
clearly dominate the errors
at small $R$. For the  subsequent analysis, we have chosen
$1 \leq \mid R_2 - R_1\mid \leq 2$.

The parametrization dependence
of $F$ is displayed in Fig.~\ref{fig:sensitivity}.
The circles therein refer
to Eq.~(\ref{potform}) and squares to the Cornell form
(Eq.~(\ref{cornell})) of the
potential. They differ by much less
than the systematic errors we state. Note that, for the purpose
of this figure, the (horizontal) shift in the tangent
point $R$ has been converted into a (vertical)
shift in $F$.
\begin{figure}[htb]
\vskip -1.1cm
\epsfxsize=310pt
\epsfbox[80 210 680 530]{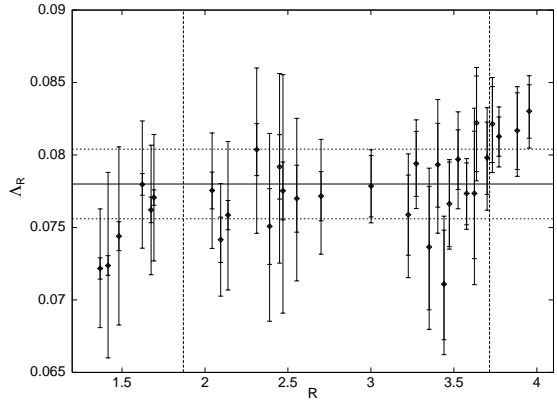}
\caption{The plateau in $\Lambda_R(R)$ at $\beta = 6.4$. The inner
(outer) part of the errorbars refers to the statistical (systematic)
uncertainties.\vskip -.3cm}
\label{fig:plateau}
\end{figure}
\section{TWO-LOOP ANALYSIS}

Given  $F$, we base our $\alpha_{q\bar q}$-analysis
on the two-loop prediction
\be
\label{twoloop}
F(R) = \\
\frac{1}{6\pi b_0}\frac{1}{R^2}\left(t
-\frac{b_1}{2b_0^2}\ln (-2 t))\right)^{-1}
\ee
with $ t = \ln (\Lambda_RR)$.
In this part of the analysis,
we exclude those  pieces of data on the force $F$
that involve $V(R=1)$, since the
unfolding of lattice artifacts is not successful on this
point, according to Fig.~\ref{fig:unfold}.
\begin{figure}[htb]
\vskip -.5cm
\epsfxsize=290pt
\epsfbox[70 210 670 600]{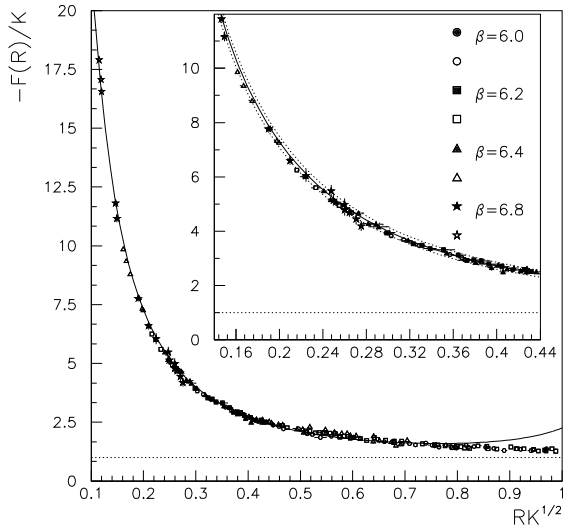}
\caption{Scaling plot of the reconstructed continuum
force.
 The perturbative
region is zoomed in the inset.
Full curve with errorband
corresponds to the two-loop formula,
 with
$\Lambda_R$ from Eq.~(12).\vskip -.3cm}
\label{fig:twoloopfit}
\end{figure}

We use Eq.~(\ref{twoloop})
to convert
the force data, $\{F(R)\}$, into a set $\{\Lambda_R(R)\}$, in order
to arrive at a sensitive test of the two-loop approximation.
A universal  plateau in
$\Lambda_R(R)$ is found in the range  $Ra \leq 1$~GeV$^{-1}$
for all $\beta$ values beyond 6.0. The situation
for $\beta = 6.4$  is  visualized in Fig.~\ref{fig:plateau}, where
the plateau is
found  to occur for $R$ values smaller than  3.7. The errors
refer both to the statistical (inner error bars) and systematic
uncertainties (outer error bars). From
the data at $\beta = 6.4$ and $ 6.8$, we obtain by
averaging
\be
\label{lambda}
\Lambda_R = .660(40) \sqrt{\sigma}\quad.
\ee
How well does the force scale? In Fig.~\ref{fig:twoloopfit}, we
have compiled all data sets with
$\beta \geq 6.0$ into one single scaling plot of the
reconstructed continuum force {\it vs.} $a$. To do that we have used
appropriate units of the string tension $\sqrt{K}$, as determined from
the potential analysis in the ``long
distance domain'', $r = Ra \geq .3$~fm.

It is gratifying to
observe that scaling is obeyed  very nicely. Moreover
the perturbative two-loop formula  describes the
$R$-dependence of the interquark force
impressively well down
 to  energies as low as $.5$~GeV, {\it i.e.}
very close to the Landau pole! We emphasize that
the $\beta = 6.5$ data points of
UKQCD~\cite{ukqcd} are in very good agreement
with these findings.
\begin{figure}[htb]
\vskip -.8cm
\epsfxsize=300pt
\epsfbox[60 210 660 530]{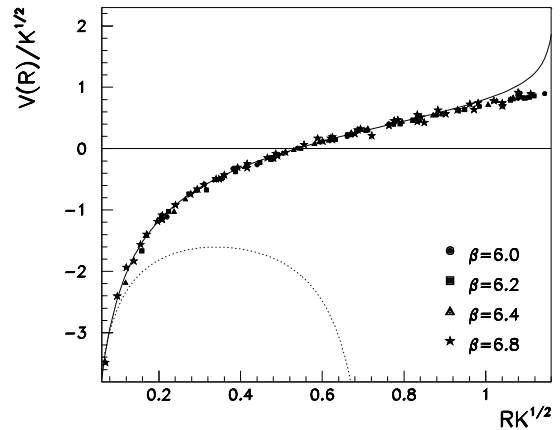}
\caption{Scaling plot of the lattice potential.
The solid curve is the integrated perturbative
force from Fig.~4; the dotted curve is $-\alpha_V(R)/R$ with
$\Lambda_V = 1.53\Lambda_R$.\vskip -.1cm}
\label{fig:potcomp}
\end{figure}

To illustrate that our results are
not biased by the analysis procedure,
we finally compare in Fig.~\ref{fig:potcomp}
the raw  lattice potential data
with  the integrated perturbative
force from Eq.~(\ref{twoloop}). We find beautiful agreement
and no indication of  a bias!

\section{SETTING THE SCALE}
It can be argued, that the string tension
is too elusive a construct and therefore
not the best ground for
laying the foundation to a scale
in lattice gauge theory.
To be precise, charmonium or  bottomium
 phenomenology is not really  sensitive to a linear
asymptotic rise of the static quark-antiquark
potential, which will die away in full QCD anyhow.
SOMMER~\cite{sommer}
proposed to base the scale fixation
on a characteristic length, $r_0$, instead, which
is connected to the intermediate
range of the potential and  therefore of  relevance to the
physical
spectrum of $\qq $-bound states. This length  obeys
the dimensionless condition
\be
R_0^2F(R_0) = \mbox{const.} = -1.65\quad .
\ee

The {\it rhs} is chosen  with an eye on phenomenological
potentials, to tune the value of the ``force radius'' $r_0 =R_0a$
close to .5 fm. From a
Cornell parametrization to the lattice potential
\be
\label{cornell}
V(R) = V_0 - e/R + KR\quad,
\ee
we compute $r_0$ and map it onto the string tension
(using the values of the fit parameters at $\beta = 6.2$), with
the result
\be
\sqrt{\sigma} \approx  1.164\, r_0^{-1}\quad .
\ee

{}From phenomenology~\cite{pheno}, we retrieve
$r_0^{-1} = 400(15)$~MeV, which induces a value
of the  redefined string tension
$\sqrt{\sigma} = 465(17)$~MeV or, with Eq.~(\ref{lambda}),
$\Lambda^{n_f=0}_{\overline{M\!S}} = 293(18)(11)$~MeV.
One can check the string tension data against the
precise data from GF11~\cite{weingarten} for the
$\rho$ mass. Apparently, we cannot get away
with quenched calculations alone, as we  find
the ratio $m_{\rho}/\sqrt{\sigma}$ to depend on $a$, at least
 in the region $\beta \leq 6.2 $.
Extrapolations allow for a continuum ratio  in the range 1.7 to 2.
For this reason, we add an  asymmetric
systematic error on the
scale which reflects both, quenching effects and scaling violations:
\be
\label{sigma}
\sqrt{\sigma} = 465^{+40}_{-100}\,\mbox{MeV}
\ee
or $\Lambda^{n_f = 0}_{\ms} = 293(18)^{+25}_{-63}$~MeV.
\section{DISCUSSION}
It is instructive to exhibit the amount of scaling violations
remnant in  different improvement schemes. This is most
apparent when plotting $\Lambda_{\ms}^{-1}$ from the two-loop
approximation {\it vs.\ } $a$.
\begin{figure}[htb]
\vskip -1cm
\epsfxsize=300pt
\epsfbox[60 210 660 530]{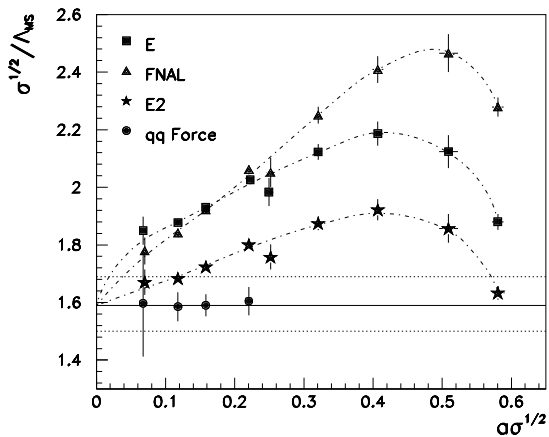}
\caption{Comparison of various `improved' coupling schemes
in terms of their apparent two-loop $\Lambda^{-1}$ values, as function
of $a$, in units of the string tension.\vskip -.3cm}
\label{fig:subasymptotia}
\end{figure}

In Fig.~\ref{fig:subasymptotia}, we compare our present results with the
estimate   from
the mean-field improved FNAL scheme~\cite{mackenzie90}
\be
\label{fnal}
 g_{\mbox{\scriptsize FNAL}}^{-2}(\pi/a) = g_0^{-2}\langle U_{\Box}\rangle
+ .025\quad .
\ee
We also
included the
results of
two other improvement schemes into Fig.~\ref{fig:subasymptotia}, denoted by
$g_E$, $g_{E^2}$.
These ``effective'' schemes
are construed to force the
perturbative series and Monte Carlo results
of the plaquette to coincide with each other
in leading and next-to-leading order,
respectively~\cite{parisi,bsrun}. The figure
provides  evidence
that $\alpha_{\qq}$ from the short range force
is  closest to the asymptotic
realm of two-loop perturbation theory!
In this sense, it looks  indeed like  a ``perfect
 coupling''!

In order to extract an estimate for $\Lambda_{\ms}$ in the
continuum limit from the improved schemes,
 one still has to perform extrapolations (see Fig.~\ref{fig:subasymptotia}).
These are
rather  delicate by their functional form, with unknown
 $a$ and $1/\ln a$ dependencies involved~\cite{protvino}.
The QCDTARO group~\cite{stamatescu93} suggested to
flatten  $\Lambda^{-1}(a)$ with  {\it brute-force},
by changing   to an ``optimized''
(yet {\it ad hoc}!!)
 renormalization
scheme through
the replacement
\be
g_{\mbox{\scriptsize FNAL}}^{-2} \lra g_{\mbox{\scriptsize fit}}^{-2} =
g_{\mbox{\scriptsize FNAL}}^{-2} + x_0\quad,
\ee
where both, $x_0$
and the three-loop coefficient in the
$\beta$-function, are treated as
free fit parameters. It turns out, however,  that
$x_0$
is unphysically large.
\begin{figure}[htb]
\vskip -1.4cm
\epsfxsize=310pt
\epsfbox[70 200 670 520]{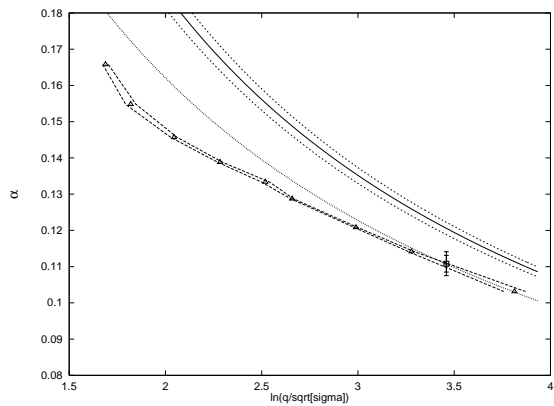}
\caption{Comparison of three lattice predictions to $\alpha_{\ms}$
for $n_f = 0$. The upper curve is from the present
analysis, Eqs.~(11,12).
$\Delta $ : FNAL-$\ms$-scheme; $\Box$: result of Ref.~[7]
with its two-loop evolution (dotted line).\vskip -.3cm}
\label{fig:alfaquenched}
\end{figure}

In Fig.~\ref{fig:alfaquenched}, we display $\alpha_{\ms}$ {\it vs.\ }
$\ln q$, with the momentum in units of the string tension
to avoid overall scale errors, in the zero flavour world.
The upper error band refers to the two-loop evolution of our
$\alpha_{\qq}$ values. Two sets of
data are shown in comparison: the triangles refer to the
estimates from Eq.~(\ref{fnal}), with $a$ determined from
the string tension,
while the point with error bar is the result of
method (c), presented by WOLFF to this conference~\cite{wolff}.
Note from Eq.~(\ref{smallcoeff}) that
our coupling scheme is much closer to the $\ms$
scheme than theirs,
as
expressed by the relation~\cite{wolff}
\be
\label{relacmd}
\alpha_{\ms} = \alpha_{C\!D\!M} + 1.256\alpha^2_{C\!D\!M}
+ {\cal O}(\alpha^3_{C\!D\!M}).
\ee
\section{UNQUENCHING}
Having established the method in the pure gauge sector, we
can now ask the sneaking question `what will happen in the real world
with dynamical fermions present?' In the absence of full fledged
QCD simulations, we have to resort to sneaky answers, {\it i.e.} rough
estimates~\cite{protvino} based
on early day full QCD results.

For that purpose, we will exploit the work of the
MT$_c$ collaboration who  performed a matching of the
quenched and unquenched (4 flavour staggered fermions)
potentials~\cite{mtc} by tuning $\beta$.
Their results enable us  to compute the
 FNAL-$\ms$ couplings, in the zero- and four- flavour
worlds.  From the latter, we can extract
the matching of the accompanying $\Lambda$-values, and
find $\Lambda_{\ms}^{(4)} = (.44 \pm .10)\Lambda_{\ms}^{(0)}$.
We assigned a $12 \%$ ($10 \%$) error due to  matching ( scaling violation)
uncertainties. The  degeneracy of the lattice  quarks
 leads us to increase the systematic error:
\be
\Lambda_{\ms}^{(4)} =
(.44^{+10}_{-14})\Lambda_{\ms}^{(0)}=129\pm8^{+43}_{-60}\,\mbox{MeV}\quad.
\ee
This can be converted into an $\alpha$-value
in the five flavour world at the $Z_0$ mass:
\be
\label{result}
\alpha_{\ms}(m_Z) = .1020^{+54}_{-104}\quad .
\ee
This lattice estimate appears to be rather low, compared to the
average value from LEP experiments, which reads
$\alpha_{\ms}^{\mbox LEP} = .118(7)$.
\section{CONCLUSION}
We have seen that a direct determination of the strong coupling from
the interquark force is feasable with the computing power of an 8K
Connection Machine CM-2.
The final number for $\alpha_s(m_Z)$, shown in Eq.~(\ref{result}),
carries errors which are comparable to the experimental ones.
At this stage, by far most of this estimated
 error is due to scale uncertainties ($15 \%$
in $\Lambda^{n_f=0}_{\ms}$, mainly from quenching)
and the cavalier $ 0 \ra 4$ flavour conversion
 ($30 \%$).

The method is applicable to full QCD.
Discretization effects can still be reduced in the present framework by
going to finer  lattices.
Future effort should also go into the direction of
a full lattice perturbative analysis.


\begin{thebibliography}{99}
\bibitem{hasenfratz}{A.\ and P.~Hasenfratz, \plb 93 (1980) 165;
\npb 193 (1981) 210.}
\bibitem{mackenzie90}{G.P.~Lepage and P.B.~Mackenzie, \cona 173;
\prd 48 (1993) 2250.}
\bibitem{michael}{Ch.~Michael, \plb 283 (1992) 103.}
\bibitem{bsrun}{G.S.~Bali and K.~Schilling, \prd 47 (1993) 661;
\conc  513.}
\bibitem{luescher1}{M.~L\"uscher et al., \npb 384 (1992) 168.}
\bibitem{luescher92}{M.~L\"uscher et al.,
\npb 389 (1993) 247; M.~L\"uscher et al., \conc 139.}
\bibitem{wolff}{U.~Wolff, this volume; M.~L\"uscher et al., DESY  preprint
DESY-93-114~(1993).}
\bibitem{elkhadra}{A.X.~El Khadra, this volume.}
\bibitem{protvino}{G.S.~Bali, Wuppertal preprint WUB 93-37.}
\bibitem{rebbibarkai}{
D.~Barkai et al., \prd 30 (1984) 1984.}
\bibitem{creutz79}{M.~Creutz, \prd 21 (1980) 2308.}
\bibitem{bspot}{G.S.~Bali and K.~Schilling, \prd 46 (1992) 2636.}
\bibitem{ukqcd}{S.P.~Booth et al., \plb 294 (1992) 385;
Ch.~Michael, \conc  509.}
\bibitem{billoire}{A.~Billoire, \plb 104 (1981) 472.}
\bibitem{rebbilang}{C.B.~Lang and C.~Rebbi, \plb 115 (1982) 137.}
\bibitem{sommer}{R.~Sommer, DESY preprint 93-062.}
\bibitem{pheno}{E.~Eichten et al., \prd 21 (1980) 203.}
\bibitem{weingarten}{F.~Butler et al, \prl 70 (1993) 2849.}
\bibitem{parisi}{F.~Karsch and R.~Petronzio, \plb 153 (1985) 87.}
\bibitem{stamatescu93}{N.~Stamatescu, this volume.}
\bibitem{mtc}{MT$_c$ Collaboration:
K.D.~Born et al., \conb 394; E.~Laermann, private communication.}
\end{thebibliography}
\end{document}